# Spatial Information and the Legibility of Urban Form: Big Data in Urban Morphology

Geoff Boeing
University of Southern California

**Abstract:** Urban planning and morphology have relied on analytical cartography and visual communication tools for centuries to illustrate spatial patterns, propose designs, compare alternatives, and engage the public. Classic urban form visualizations – from Giambattista Nolli's ichnographic maps of Rome to Allan Jacobs's figure-ground diagrams of city streets – have compressed physical urban complexity into easily comprehensible information artifacts. Today we can enhance these traditional workflows through the Smart Cities paradigm of understanding cities via user-generated content and harvested data in an information management context. New spatial technology platforms and big data offer new lenses to understand, evaluate, monitor, and manage urban form and evolution. This paper builds on the theoretical framework of visual cultures in urban planning and morphology to introduce and situate computational data science processes for exploring urban fabric patterns and spatial order. It demonstrates these workflows with OSMnx and data from OpenStreetMap, a collaborative spatial information system and mapping platform, to examine street network patterns, orientations, and configurations in different study sites around the world, considering what these reveal about the urban fabric. The age of ubiquitous urban data and computational toolkits opens up a new era of worldwide urban form analysis from integrated quantitative and qualitative perspectives.

## 1. Introduction

Information management is an important component of urban research and praxis. Data-driven modeling and exploration of cities entail cycles of information management activities – from data acquisition and transformation, to structuring and interpreting data to produce useful information, to knowledge dissemination for planning and advocacy. This paper considers urban morphology through these information management processes in the Smart Cities paradigm.

---

The Smart Cities paradigm aims to understand cities through user-generated content, ubiquitous sensing, and automatically harvested data (Batty et al., 2012). While this often takes the form of a modernized cybernetics through top-down monitoring, optimization, and control, there is greater potential in this paradigm's critical turn to richer problem sets (Goodspeed, 2015; Kitchin, 2016). Rather than considering livability as an optimization problem, planners might use urban data to enrich socio-political processes of community advocacy, understanding, consensus-forming, and public decision-making. Critical approaches might identify sampling biases in user-generated content to adjust for over-representation of certain groups in these datasets and to foreground marginalized voices. User-generated spatial data can be used to introspectively unpack planning and design histories and the spatial logics they have manifested in different places as a function of era, politics, culture, and local economic conditions. Interpretative and narrative approaches can enrich and contextualize data-driven urban morphology (Erin et al., 2017; Moudon, 1997).

Spatial information (and in turn information management) plays a central role in this space as nearly all urban and human processes are spatially-situated. In particular, this paper reflects on user-contributed big data about spatial infrastructure that allow us to examine the physical substrate that constrains and shapes the flows of people, goods, and information through urban space (Batty, 2013). Street networks are the paradigmatic example of such infrastructure. Researchers have explored these networks in recent years to model trips and traffic, to uncover fundamental patterns in city organization, and to explore urban planning and design histories. OpenStreetMap (OSM) – a worldwide mapping community and online geospatial information system – has played an increasingly key role in these streams of scholarship as it provides a freely available, high quality source of data on street networks and other urban infrastructure worldwide (Barrington-Leigh and Millard-Ball, 2017; Jokar Arsanjani et al., 2015).

This paper builds on recent work by Crooks et al. (2016) – who argued that big data is an underexplored, emerging frontier in urban morphology – to introduce and review examples of analyses that help to reveal and explain the urban form. In particular, it focuses on a motivating question: how can user-contributed big data be collected, organized, and modeled to explore the various spatial logics resulting from urban planning, design, and evolution? It takes up Crooks et al.'s call to develop workflows that integrate both data-driven and qualitative (i.e., interpretive or narrative) methods of urban morphology in today's era of ubiquitous urban big data. Situating this theoretical work in the visual culture of planning, this paper presents a visualization-mediated interpretative process of urban morphology.

This paper is organized as follows. First it situates analytical cartography within the theoretical framework of the visual culture of planning, which has traditionally sought to visualize urban space to understand city presents and futures (Gage, 2009; Gissen, 2008; Jacobs, 1984; Shanken, 2018; Söderström, 1996; Tobler, 1976) – from Nolli maps (Hwang and Koile, 2005; Verstegen and Ceen, 2013) to figure-ground street diagrams (Jacobs, 1995) to rose diagrams (Mohajeri and Gudmundsson, 2014). Then it discusses an information management workflow for collecting, modeling, analyzing, and visualizing OSM big data using open source tools like OSMnx and computational methodologies. Finally, it considers what these data-driven techniques reveal about different places and modes of urbanization before discussing implications for the Smart Cities paradigm of understanding cities. It argues that such historically- and culturally-informed quantitative methods are essential for understanding the patterns and forms resulting from urban processes.

## 2. Spatial Information and Urban Form

Much like the broader field of information management, many of the former central challenges in collecting, storing, and sharing spatial data have now evolved into commonplace processes and standardized platforms to largely alleviate technical burden. Today, greater emphasis belongs to the role that spatial information systems play in shaping urban patterns and processes from travel behavior, to housing markets and gentrification, to the ways in which planners, designers, and citizens engage with spatial information (Batty, 2005; Evans-Cowley and Griffin, 2012). This engagement is critical both for evidence-based city planning as well as for collaborative community-building in an era of ubiquitous technology. In particular, the Smart Cities paradigm today aims to quantify and measure urban patterns and processes in a positivist approach to understanding, controlling, and improving cities through information technology (Angelo and Vormann, 2018; Kitchin, 2016; Krivý, 2018; Townsend, 2015; Watson, 2015). As an epistemological project, Smart Cities research and practice take many diverse forms (Albino et al., 2015; Ismagilova et al., 2019; Lytras and Visvizi, 2018; Stone et al., 2018; Visvizi et al., 2018).

One such subject of urban morphology research today intersects Smart Cities, computational geometry, and network science to explore the patterns of urban form and circulation through large, harvested, user-generated data sets. Scholars have investigated this to explore travel behavior, public health and safety, and residential sorting (Ewing and Cervero, 2010; Hajrasouliha and Yin, 2015; Levinson and El-Geneidy, 2009; Marshall et al., 2014; Porta et al., 2012; Southworth and Ben-Joseph, 1995; Xiao et al., 2016; Zhong et al., 2014). Spatial information plays an important role in urban simulation for community visioning (Vanegas et al., 2009; Waddell et al., 2018), streetscape quality analysis (Shen et al., 2018), and predicting urban attributes from street imagery (Arietta et al., 2014). Another research stream examines the spatial ordering of cities' configuration and orientation through circulation network patterns (Barthelemy et al., 2013; Barthelemy, 2017; Boeing, 2019; Buhl et al., 2006; Chan et al., 2011; Courtat et al., 2011; Gudmundsson and Mohajeri, 2013; Louf and Barthelemy, 2014; Mohajeri et al., 2013b; Mohajeri and Gudmundsson, 2012, 2014). Various spatial logics and ordering principles exist in planned, unplanned, formal, informal, gridded, and organic urban patterns (Kostof, 1991; Rose-Redwood and Bigon, 2018; Smith, 2007) – and a city without one single, formal, geometric ordering logic may have well-defined, high-functioning physical and social structure (Hanson, 1989).

Data-driven urban morphology seeks to explore urban form and spatial order – modeling spatial data to trace histories, configurations, and orientations in physical space – but requires a grounding in narrative and interpretative approaches to reveal the nuance of local context and history. The urban historian Spiro Kostof (1991, p. 10) said: "Form, in itself, is very lamely informative of intention. We 'read' form correctly only to the extent that we are familiar with the precise cultural conditions that generated it… The more we know about cultures, about the structure of society in various periods of history in different parts of the world, the better we are able to read their built environment." That is, urban spatial data must be contextually interpreted to become meaningful information. Kostof (*ibid.*, p. 11) continues: "There is no point in noticing the formal similarities between L'Enfant's plan for Washington and the absolutist diagrams of Versailles or Karlsruhe… unless we can elaborate on the nature of the content that was to be housed within each, and the social premises of the designers."

Here we specifically situate such visual thinking within the theoretical frameworks developed by Söderström (1996), Gissen (2008), and Shanken (2018). Söderström traces the distinct histories of cartography, geography, and postmodernity, arguing that the visual representation of urban form undergirds the scientific mode of studying cities. This visual thinking renders abstract spatial information legible to planners and citizens. Gissen explores the historical and still evolving visual-analytical links between architecture and geography, arguing that datascapes and visualization mediate the differences between empirical research traditions and the creative process of urban design. Shanken argues that urban planners have employed a constellation of visual methods to analyze spatial information and represent the city. This representational visual culture was historically exemplified by Giambattista Nolli's eighteenth century ichnographic study of Rome, producing the famous figure-ground Nolli Maps of the urban fabric (Hwang and Koile, 2005; Verstegen and Ceen, 2013). Two centuries later, this particular methodology was explored anew by Allan Jacobs' (1995) comparative visual study of dozens of urban street networks around the world.

This paper takes up where these traditional manual workflows left off to explore new computational, automated, big data methods through three demonstrative questions important to urban morphology research. 1) How do street networks embody specific spatial logics to organize city dynamics? 2) How do their orientations and configurations vary across places accordingly? 3) How can we visualize these complex patterns to render their information legible to planners and citizens using modern computational workflows and spatial big data?

### 3. Working with OpenStreetMap Data

This paper explores these questions using case studies in a small set of cities and building models of urban street networks from OSM raw data using OSMnx, a Python package to automate and streamline OSM spatial data acquisition and analysis (Boeing, 2017). Alone, OSM's spatial information system and its various APIs can be challenging to synthesize and work with in nuanced and theoretically-sound ways (particularly for planning practitioners): its raw data do not lend themselves automatically to urban form/network analysis and its custom query languages can be cumbersome for scripting. OSMnx allows researchers and practitioners to easily download street network, building, and amenity data for any study site in the world, then automatically construct them into street network graphs or spatial data frames for built-in visualization and statistical analysis. This empowers new ways of engaging with this massive repository of global spatial information.

These models are nonplanar directed multigraphs, where graph nodes represent intersections and dead-ends, and graph edges represent the street segments linking them (Barnes and Harary, 1983; Cardillo et al., 2006; Lin and Ban, 2013; Marshall et al., 2018; Trudeau, 1994). OSMnx allows users to download spatial data from OSM for any study site boundary in the world, automatically construct it into a model that conforms to urban design and transportation planning conventions, and then analyze and visualize it (Boeing, 2017). OSM is a valuable source of geospatial data as it has worldwide coverage, generally high quality, and an active collaborative user community (Barron et al., 2014; Basiri et al., 2016; Corcoran et al., 2013; Girres and Touya, 2010; Haklay, 2010; Jokar Arsanjani et al., 2015; Neis et al., 2011; Over et al., 2010; Zielstra et al., 2013).

This paper employs two visualization methods to illustrate urban morphology through street patterns. First, it uses OSMnx to produce figure-ground diagrams of street networks and building footprints to compress urban form complexity, illustrate urban design, and communicate planning decisions and histories. These are inspired by Allan Jacobs's (1995) classic book on street-level urban form and design, which featured dozens of hand-drawn figure-ground diagrams in the general style of Nolli maps, depicting built versus open space to illustrate street network patterns. We adapt this cartographic methodology to a computational, big data workflow to likewise depict one square mile of multiple cities' street networks, considering similarities and differences. Plotting cities at the same scale provides a revealing spatial objectivity in visually comparing their street networks and urban forms.

The second method uses rose diagrams of street orientations to visualize the spatial ordering of urban circulation infrastructure (Gudmundsson and Mohajeri, 2013; Mohajeri et al., 2013a, 2013b; Mohajeri and Gudmundsson, 2014, 2012). First we calculate the compass bearings of all the street segments in 25 world cities, then visualize them with a rose diagram in which the bars' directions represent 10° bins around the compass, and the bars' lengths represent the relative frequency of street segments that fall in each bin (Boeing, 2019). This produces a visual representation of the extent to which a street network follows the spatial ordering of a low entropy grid versus having streets oriented evenly in all compass directions.

## 4. Urban Circulation Systems and Spatial Logics

Figure 1 shows one square mile figure-ground diagrams from 12 cities around the world. At the top-left, Portland, Oregon and San Francisco, California typify the late nineteenth century

American orthogonal grid (Cole, 2014; Marshall et al., 2015; Southworth and Ben-Joseph, 1997, 1995). Portland's famously compact, walkable, 200-foot × 200-foot blocks are clearly visible but its grid is interrupted by the Interstate 405 freeway which tore through the central city in the 1960s (Mesh, 2014; Speck, 2012). In the middle-left, the business park in suburban Irvine, California demonstrates the coarse-grained, modernist, auto-centric form that characterized American urbanization in the latter half of the twentieth century (Hayden, 2004; Jackson, 1985; Jacobs, 1995).

In stark contrast, Rome has a more fine-grained, complex, organic form which evolved over millennia of self-organization and urban planning (Taylor et al., 2016). Representing each of these street networks here at the same scale – one square mile – it is easy to compare the qualitative urban patterns in these different cities to one another. Contrast the order of the nineteenth century orthogonal grid in San Francisco and the functionalist simplifications of twentieth century Irvine to the messy, complex mesh of pedestrian paths, passageways, and alleys constituting the circulation network in the ancient center of Rome.

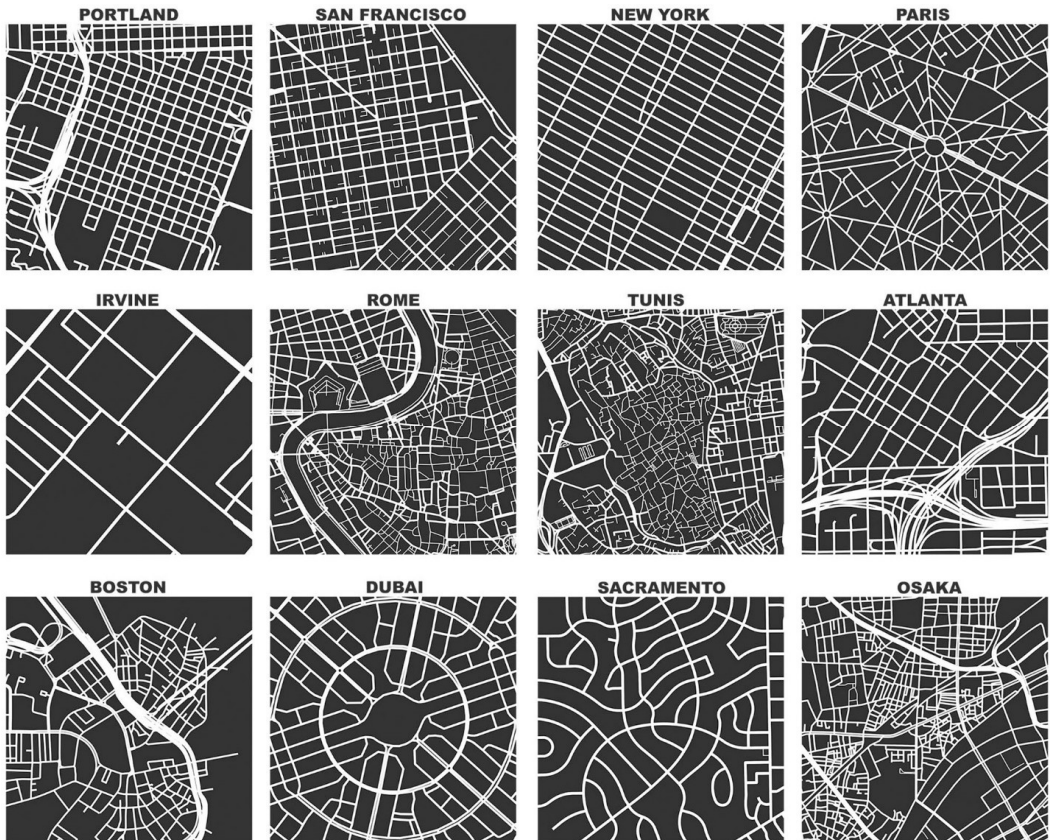

**Figure 1.** One square mile of each city's street network. The consistent spatial scale allows us to easily compare different kinds of street networks and urban forms in different kinds of places.

At the top- and middle-right, we see New York, Paris, Tunis, and Atlanta. Midtown Manhattan's rectangular grid originates from the New York Commissioners' Plan of 1811, which laid out its iconic 800-foot × 200-foot blocks approximately 29 degrees off true North (Ballon, 2012; Koeppel, 2015). Broadway weaves diagonally across it, revealing the path dependence of the old Wickquasgeck Trail's vestiges, which Native American residents used to traverse the length of the island long before the first Dutch settlers arrived (Holloway, 2013; Shorto, 2004).

At the center of the Paris square mile lies the Arc de Triomphe, from which Baron Haussmann's streets radiate outward as remnants of his massive demolition and renovation of nineteenth century Paris (Hall, 1996). The spatial signatures of Haussmann's project can clearly be seen via network analysis through the redistribution of betweenness centralities and block sizes (Barthelemy et al., 2013). At the center of the Tunis square mile lies its Medina, with a complex urban fabric that evolved over the middle ages (Kostof, 1991; Micaud, 1978). Finally, Atlanta is typical of many American downtowns: coarse-grained, disconnected, and surrounded by freeways (Allen, 1996; Grable, 1979; Jackson, 1985; Kruse, 2007; Rose, 2001).

The bottom row of Figure 1 shows square miles of Boston, Dubai, Sacramento, and Osaka. The central Boston square mile includes the city's old North End – beloved by Jane Jacobs (1961) for its lively streets, but previously cut-off from the rest of the city by the Interstate 93 freeway. This freeway has since been undergrounded as part of the "Big Dig" megaproject to alleviate traffic and re-knit the surface-level urban fabric (Flyvbjerg, 2007; Robinson, 2008). The Dubai square mile shows Jumeirah Village Circle, a master-planned residential suburb designed in the late 2000s by the Nakheel corporation, a major Dubai real estate developer (Boleat, 2005; Haine, 2013; Kubat et al., 2009). Its street network demonstrates a hybrid of the whimsical curvilinearity of the garden cities movement and the ordered geometry of modernism. The Sacramento square mile depicts its northeastern residential suburb of Arden-Arcade and demonstrates Southworth and Ben-Joseph's (1997) "warped parallel" and "loops and lollipops" design patterns of late twentieth century American urban form.

Finally, the Osaka square mile portrays Fukushima-ku, a mixed-use but primarily residential neighborhood first urbanized during the late nineteenth century. Today, the freeway we see in the upper-right of this square mile infamously passes through the center of the high-rise Gate Tower Building's fifth through seventh floors (Yakunicheva, 2014). This peculiar intermingling of street network and edifice arose when transportation planners were forced to compromise with private landowners seeking to redevelop their property, despite the prior designation of the freeway's alignment (Isaac, 2014).

To qualitatively compare urban spatial forms in different kinds of places, these visualizations depict a mix of modern central business districts, ancient historic quarters, twentieth century business parks, and suburban residential neighborhoods. The cities they represent are drawn from across the United States, Europe, North Africa, the Arabian Peninsula, and East Asia. Yet street network patterns also vary greatly within cities: Portland's suburban east and west sides look different than its downtown, and Sacramento's compact, grid-like downtown looks different than its residential suburbs – a finding true of many American cities (Boeing, 2018a). A single square mile diagram thus cannot be taken to be representative of broader spatial scales or other locations within the municipality. These visualizations, rather, show us how different urbanization patterns and paradigms compare at the same scale, using automatically harvested user-generated data. This can serve both as a practitioner's tool for comprehending the physical outcomes of planning and informal urbanization, as well as a tool for communicating urban planning and design in a clear and immediate manner to laymen – leveraging spatial information to improve political collaboration and multi-level co-governance.

These uses can be seen perhaps even more clearly when we use OSMnx to visualize street networks along with OSM building footprints, as shown in Figure 2. At the top-left, we see the densely built form of midtown Manhattan, with large buildings filling most of the available space between streets. Within this square mile, there are 2,237 building footprints with a median area of 241 square meters. At the top-right, we see the medium-density perimeter blocks of San Francisco's Richmond district, just south of the Presidio. Here the building footprints line the streets while leaving the centers of each block as open space for residents. Within this square mile, there are 5,054 building footprints with a median area of 142 square meters. The bottom two images in Figure 2 reveal an entirely different mode of urbanization by visualizing the slums of Monrovia, Liberia and Port-au-Prince, Haiti. These informal settlements are much finer-grained and are not structured according to the orderly, centralized planning of the American street grids in the top row. Monrovia's square mile contains 2,543 building footprints with a median area of 127 square meters. Port-au-Prince's square mile contains 14,037 building footprints with a median area of just 34 square meters.

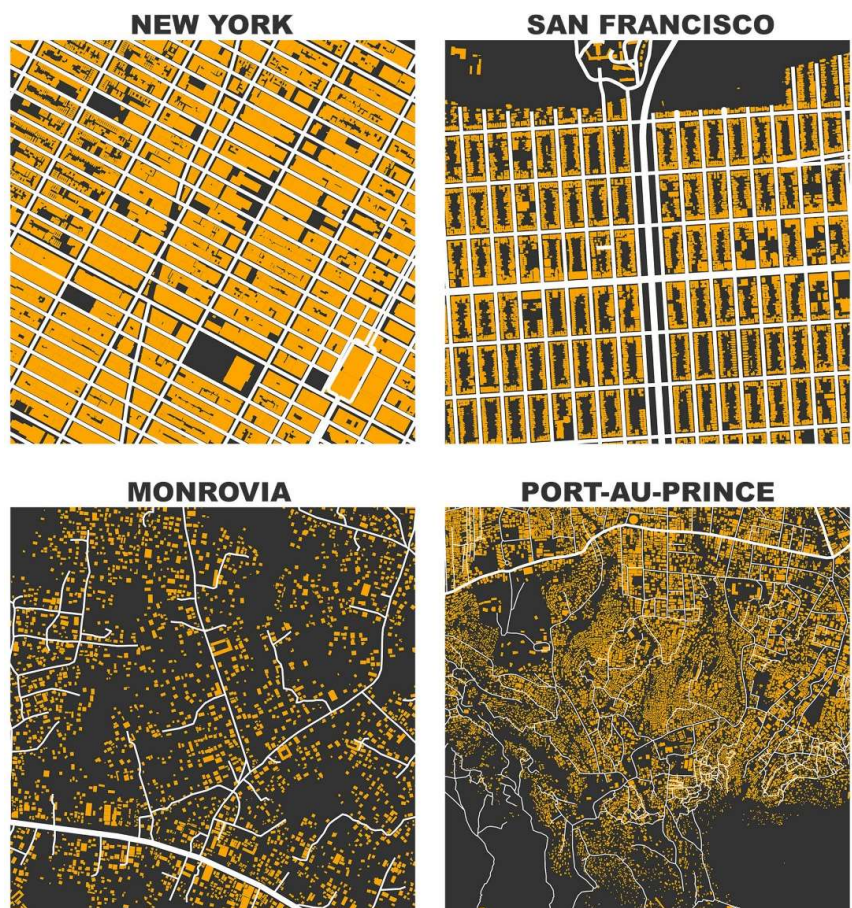

**Figure 2.** One square mile of each city's street network and building footprints, comparing US cities to informal settlements in the Global South.

OSM data and OSMnx provide planning practitioners an easy-to-use tool to visualize and examine street networks and building footprints as a planning and communication tool. The data in Figure 2, for instance, could help planners and residents in Monrovia and Port-au-Prince collaboratively study how to percolate formal circulation networks into these informal settlements with minimal disruption to the existing urban fabric, homes, and livelihoods (Brelsford et al., 2019, 2018, 2015; Masucci et al., 2013; Zook et al., 2010).

Visualizing spatial information can also reveal the state assertion of power and modernism's inversion of traditional urban spatial order (Holston, 1989; Vale, 2008). In pre-industrial cities, the figure dominates the ground as the diagram displays scattered open space between buildings, as seen in Figure 3. But in modernist cities, the ground dominates the figure as only a few scattered buildings are positioned as sculptural elements across the landscape's void. The modernist paradigm sought to open up the dense, messy, and complex urban fabric with towers-in-the-park, spacing, highways, and functional simplicity (Boeing, 2018b; Fishman, 2011). This phenomenon is clearly seen in Brasília, the modernist capital of Brazil, designed as a planned city in the 1950s by Lúcio Costa, Oscar Niemeyer, and Roberto Burle Marx (Figure 3). The structural order of the city also suggests "an ordering of social relations and practices in the city" (Holston, 1989, p. 125). These figure-ground diagrams provide a spatial data-driven method to in turn qualitatively interpret and study the urban form and circulation networks that structure human activities and social relations.

The rose diagrams in Figure 4 offer another perspective on visualizing this structural ordering of the city. Each polar histogram visualizes the orientation (compass bearing) of the borough's street segments, with bins representing 10-degrees around the compass and bar lengths representing relative frequency (for complete methodological details and theoretical development see Boeing, 2019). For example, in Manhattan's rose diagram we can see the spatial order produced by its dominant orthogonal grid (cf. Figure 1) as its street bearings are primarily captured in four bins, offset from true North. The other boroughs have higher entropy street orientations, not adhering as strictly to the ordering logic of a single grid. Such an analysis can be performed with tools like QGIS or ArcGIS, but OSMnx helps to streamline the workflow by automating the process of downloading the raw data, building the model, and then allowing easy directionality analysis using Python libraries like geopandas and matplotlib and the interactivity afforded by the Jupyter environment.

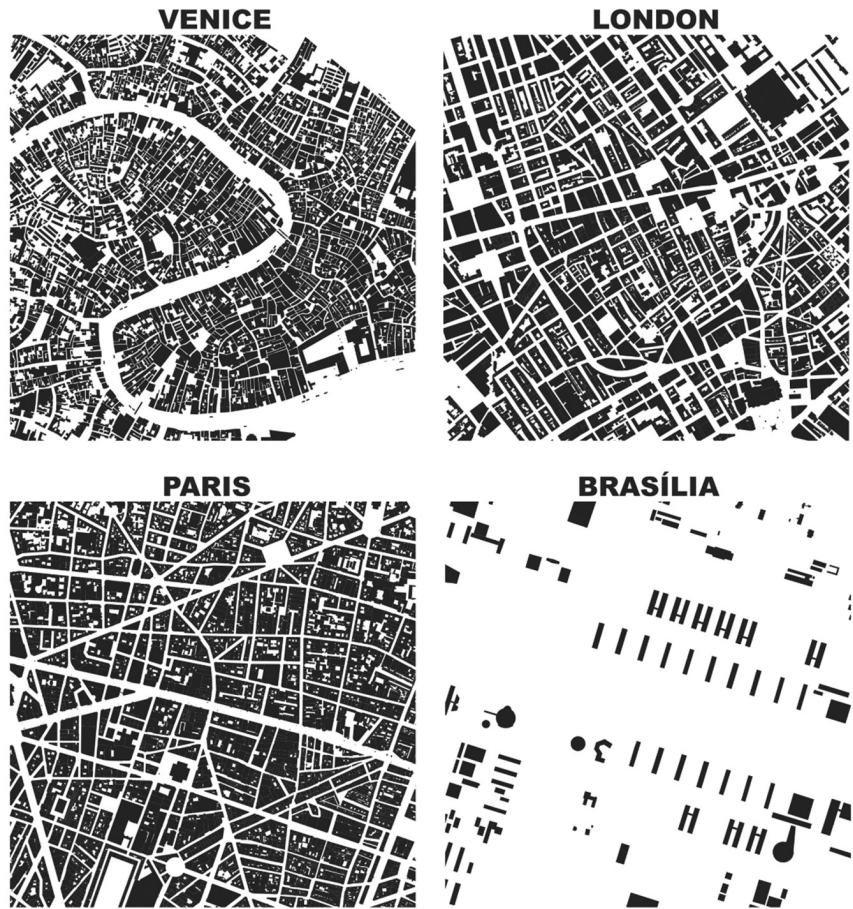

**Figure 3.** One square mile figure-ground diagrams of building footprints in the centers of Venice, London, Paris, and Brasília reveal the modernist inversion of traditional urban spatial order.

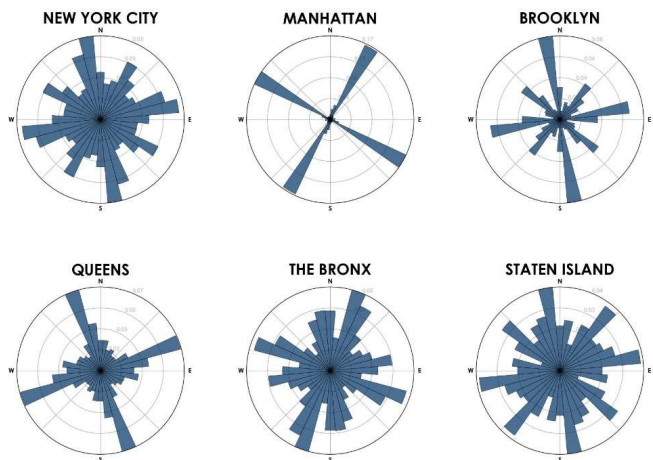

**Figure 4.** Rose diagrams of the street orientations in New York City and its five constituent boroughs.

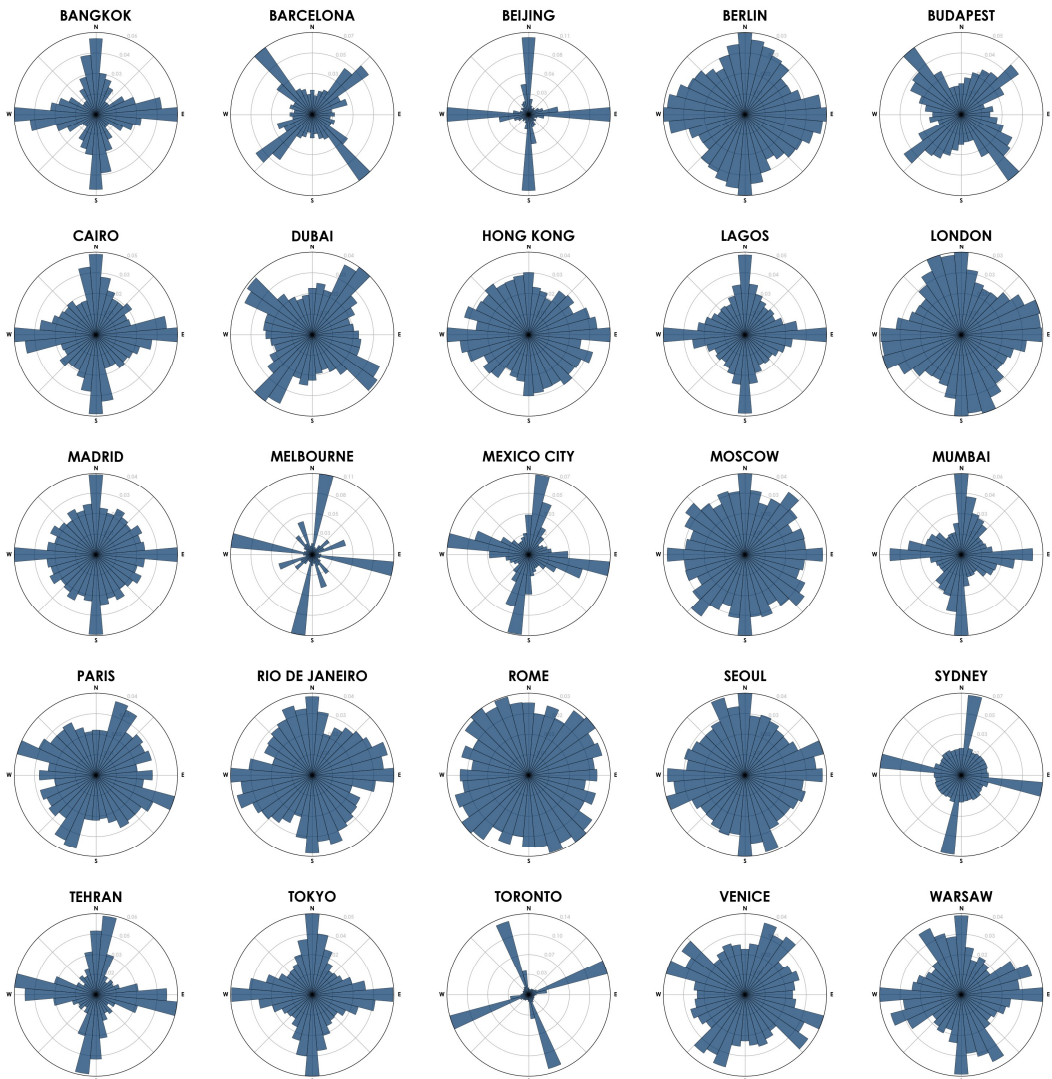

**Figure 5.** Rose diagrams of the street orientations in 25 cities around the world.

In Figure 5 we see rose diagrams of 25 cities (municipalities) around the world. This study scale aggregates heterogeneous neighborhoods into a single analytical whole, but offers the benefit of capturing the scale of city planning jurisdiction to tell us about the spatial ordering that the circulatory system provides. While some street networks in modern cities in Canada, Australia, and China demonstrate similar low-entropy grids, far more of these cities show higher entropy. That is, their streets are oriented more evenly in all compass directions rather than following the spatial ordering logic of one or two consistent grids. The spatial signature of the street grid is clearest in cities like Toronto and Beijing, while cities like Rome and Rio de Janeiro demonstrate more-organic and less-orthogonal patterns. The patterns in Beijing are interesting as they deviate from many of its Asian neighbors, instead conforming more to the rationalist, centrally planned gridirons of Western cities like Toronto, Melbourne, and Manhattan and suggesting a certain spatial logic undergirding its massive and rapid urbanization in recent years.

## 5. Discussion

Through the tools of urban morphology and computer science, spatial information allows us to see how urban planning, design, and millions of individual decisions shape how cities organize and order space according to various spatial social logics and cultures. In the 1990s, Allan Jacobs adapted the style of Nolli maps to manually illustrate figure-ground diagrams of cities to explore their urban form. Today, OSM provides a worldwide data set to perform these analytical cartographic workflows automatically and computationally in the Smart Cities mode of analyzing and understanding cities through user-generated big data.

This paper operationalized OSMnx's figure-ground diagrams and rose diagrams as methods of hybrid quantitative-qualitative analysis of urban patterns. These visualizations reveal the texture, grain, and spatial-ordering logic of different cities around the world. Compressing the dense spatial complexity inherent in cities, they offer a streamlined and legible window into the urban fabric and how the circulation system's infrastructure percolates through it. These figure-ground diagrams allow us to compare across places at the same scale to better understand similarities and differences particularly in texture, grain, and connectivity. The rose diagrams compress the complexity of street network orientation entropy into simple plots that immediately reveal the spatial ordering of the city's streets and its underlying spatial logic.

These visual urban morphology methods and OSMnx workflows can help planners convey comparative urban form to laypersons. They can destigmatize density and explain how connectivity and urban texture vary across cities. They simplify complicated urban planning and urban data science concepts to make them more approachable for practitioners, community advocates, and other members of the public to engage in citizen science. Future research can further operationalize these techniques to quantify circulation system connectivity and griddedness, and measure how street network design has evolved across the United States. This can provide a comprehensive understanding of a city's morphological trajectory through time and, in turn, help planners collaboratively shape that trajectory.

## 6. Conclusion

This paper explored information management processes of data acquisition, transformation, modeling, interpretation, and dissemination for understanding and communicating urban form. It took up Crooks et al.'s (2016) call for developing workflows that meld both data-driven and interpretive approaches to urban morphology in the burgeoning era of user-generated big data. It discussed the OSMnx toolkit for modeling, analyzing, and visualizing street networks from morphological perspectives, advancing techniques such as Jacobs's (1995) figure-ground diagrams and rose diagrams into reproducible, computational workflows. It situated this theoretical work in the visual culture of planning – adopting a visually-mediated narrative approach of exploring the urban form and in particular demonstrating how OSM data can be modeled to explore the spatial ordering that results from urban planning and design.

In doing so, it made three primary contributions to urban planning/morphology at the intersection of Smart Cities and information management. First, it presented an urban data science workflow for collecting, modeling, and interpreting user-contributed big data to research how cities' street networks embody different spatial logics and organize complex human dynamics through urban space. Second, it demonstrated a big data-driven approach to modeling and evaluating street network orientation and configuration in different places. Third, it extended these quantitative streams of Smart Cities/information management scholarship in a complementary qualitative direction by describing how planners can visualize these complex patterns to make them legible to citizens in collaborative planning processes of comparing alternative scenarios. This communicative data-driven decision-making sits at the heart of the intersection of urban planning, morphology, and information management.

Throughout, this paper operationalized a narrative approach, following Kostof's (1991) emphasis on social premises and cultural conditions, to interpret urban morphology qualitatively through data-driven quantitative analysis and visualization. This sets up two key next steps for future urban morphology research in the Smart Cities paradigm. First, from a big data perspective, developing richer interactive visual functionality is critical to empowering analytical reasoning in these planning and engagement processes. Some interaction was discussed in this paper, but this remains an important emerging frontier at the intersection of information management and urban planning. Second, future research at the intersection of information science and urban morphology can further develop the rose diagrams in the context of information entropy to explore street network orientation-order and develop new indicators of griddedness to see how planning and design paradigms have evolved over time in different cities and countries. In tandem, information management and urban morphology will further converge to provide comprehensive understandings of city pasts and presents and empower planners and community members in collaborative data-driven decision-making processes.

## References

Albino, V., Berardi, U., Dangelico, R.M., 2015. Smart Cities: Definitions, Dimensions, Performance, and Initiatives. Journal of Urban Technology 22, 3–21. https://doi.org/10.1080/10630732.2014.942092

Allen, F., 1996. Atlanta Rising: The Invention of an International City 1946-1996. Taylor Trade Publishing, Atlanta, GA.

Angelo, H., Vormann, B., 2018. Long waves of urban reform: Putting the smart city in its place. City 22, 782–800. https://doi.org/10.1080/13604813.2018.1549850

Arietta, S.M., Efros, A.A., Ramamoorthi, R., Agrawala, M., 2014. City Forensics: Using Visual Elements to Predict Non-Visual City Attributes. IEEE Trans. Visual. Comput. Graphics 20, 2624–2633. https://doi.org/10.1109/TVCG.2014.2346446

Ballon, H. (Ed.), 2012. The Greatest Grid: The Master Plan of Manhattan, 1811-2011. Columbia University Press, New York, NY.

Barnes, J.A., Harary, F., 1983. Graph Theory in Network Analysis. Social Networks 5, 235–244. https://doi.org/10.1016/0378-8733(83)90026-6

Barrington-Leigh, C., Millard-Ball, A., 2017. The world's user-generated road map is more than 80% complete. PLOS ONE 12, e0180698. https://doi.org/10.1371/journal.pone.0180698

Barron, C., Neis, P., Zipf, A., 2014. A Comprehensive Framework for Intrinsic OpenStreetMap Quality Analysis. Transactions in GIS 18, 877–895. https://doi.org/10.1111/tgis.12073

Barthelemy, M., 2017. From paths to blocks: New measures for street patterns. Environment and Planning B: Urban Analytics and City Science 44, 256–271. https://doi.org/10.1177/0265813515599982

Barthelemy, M., Bordin, P., Berestycki, H., Gribaudi, M., 2013. Self-organization versus top-down planning in the evolution of a city. Scientific Reports 3. https://doi.org/10.1038/srep02153

Basiri, A., Jackson, M., Amirian, P., Pourabdollah, A., Sester, M., Winstanley, A., Moore, T., Zhang, L., 2016. Quality assessment of OpenStreetMap data using trajectory mining. Geospatial Information Science 19, 56–68. https://doi.org/10.1080/10095020.2016.1151213

Batty, M., 2013. The New Science of Cities. MIT Press, Cambridge, MA.

Batty, M., 2005. Network geography: Relations, interactions, scaling and spatial processes in GIS, in: Unwin, D.J., Fisher, P. (Eds.), Re-Presenting GIS. John Wiley & Sons, Chichester, England, pp. 149–170.

Batty, M., Axhausen, K.W., Giannotti, F., Pozdnoukhov, A., Bazzani, A., Wachowicz, M., Ouzounis, G., Portugali, Y., 2012. Smart Cities of the Future. The European Physical Journal Special Topics 214, 481–518. https://doi.org/10.1140/epjst/e2012-01703-3

Boeing, G., 2019. Urban Spatial Order: Street Network Orientation, Configuration, and Entropy. Applied Network Science 4, 67. https://doi.org/10.1007/s41109-019-0189-1

Boeing, G., 2018a. A Multi-Scale Analysis of 27,000 Urban Street Networks: Every US City, Town, Urbanized Area, and Zillow Neighborhood. Environment and Planning B: Urban Analytics and City Science, published online before print. https://doi.org/10.1177/2399808318784595

Boeing, G., 2018b. Measuring the Complexity of Urban Form and Design. Urban Design International 23, 281–292. https://doi.org/10.1057/s41289-018-0072-1

Boeing, G., 2017. OSMnx: New Methods for Acquiring, Constructing, Analyzing, and Visualizing Complex Street Networks. Computers, Environment and Urban Systems 65, 126–139. https://doi.org/10.1016/j.compenvurbsys.2017.05.004

Boleat, M., 2005. Housing Finance in the United Arab Emirates. Housing Finance International 19, 3.

Brelsford, C., Martin, T., Bettencourt, L.M., 2019. Optimal reblocking as a practical tool for neighborhood development. Environment and Planning B: Urban Analytics and City Science 46, 303–321. https://doi.org/10.1177/2399808317712715

Brelsford, C., Martin, T., Hand, J., Bettencourt, L., 2015. The Topology of Cities (Working Paper No. 15- 06–021). The Santa Fe Institute, Santa Fe, NM.

Brelsford, C., Martin, T., Hand, J., Bettencourt, L.M.A., 2018. Toward cities without slums: Topology and the spatial evolution of neighborhoods. Sci. Adv. 4, eaar4644. https://doi.org/10.1126/sciadv.aar4644

Buhl, J., Gautrais, J., Reeves, N., Solé, R.V., Valverde, S., Kuntz, P., Theraulaz, G., 2006. Topological patterns in street networks of self-organized urban settlements. The European Physical Journal B: Condensed Matter and Complex Systems 49, 513–522. https://doi.org/10.1140/epjb/e2006-00085-1

Cardillo, A., Scellato, S., Latora, V., Porta, S., 2006. Structural properties of planar graphs of urban street patterns. Physical Review E 73. https://doi.org/10.1103/PhysRevE.73.066107

Chan, S.H.Y., Donner, R.V., Lämmer, S., 2011. Urban road networks — spatial networks with universal geometric features? The European Physical Journal B: Condensed Matter and Complex Systems 84, 563–577. https://doi.org/10.1140/epjb/e2011-10889-3

Cole, T., 2014. A Short History of San Francisco, 3rd ed. Heyday, Berkeley, CA.

Corcoran, P., Mooney, P., Bertolotto, M., 2013. Analysing the growth of OpenStreetMap networks. Spatial Statistics 3, 21–32. https://doi.org/10.1016/j.spasta.2013.01.002

Courtat, T., Gloaguen, C., Douady, S., 2011. Mathematics and morphogenesis of cities: A geometrical approach. Physical Review E 83, 1–12. https://doi.org/10.1103/PhysRevE.83.036106

Crooks, A.T., Croitoru, A., Jenkins, A., Mahabir, R., Agouris, P., Stefanidis, A., 2016. User-Generated Big Data and Urban Morphology. Built Environment 42, 396–414. https://doi.org/10.2148/benv.42.3.396

Erin, I., Araldi, A., Fusco, G., Cubukcu, E., 2017. Quantitative Methods of Urban Morphology in Urban Design and Environmental Psychology. Presented at the 24th ISUF 2017 - City and Territory in the Globalization Age. https://doi.org/10.4995/ISUF2017.2017.5732

Evans-Cowley, J., Griffin, G., 2012. Microparticipation with Social Media for Community Engagement in Transportation Planning. Transportation Research Record 2307, 90–98. https://doi.org/10.3141/2307-10

Ewing, R., Cervero, R., 2010. Travel and the Built Environment: A Meta-Analysis. Journal of the American Planning Association 76, 265–294.
https://doi.org/10.1080/01944361003766766

Fishman, R., 2011. The Open and the Enclosed: Shifting Paradigms in Modern Urban Design, in: Banerjee, T., Loukaitou-Sideris, A. (Eds.), Companion to Urban Design. Routledge, London, England, pp. 30–40.

Flyvbjerg, B., 2007. Policy and planning for large-infrastructure projects: problems, causes, cures. Environment and Planning B: Planning and Design 34, 578–597. https://doi.org/10.1068/b32111

Gage, M.F., 2009. In Defense of Design. Log 16, 39–45.

Girres, J.-F., Touya, G., 2010. Quality Assessment of the French OpenStreetMap Dataset. Transactions in GIS 14, 435–459. https://doi.org/10.1111/j.1467-9671.2010.01203.x

Gissen, D., 2008. Architecture's Geographic Turns. Log 12, 59–67.

Goodspeed, R., 2015. Smart Cities: Moving beyond Urban Cybernetics to Tackle Wicked Problems. Cambridge Journal of Regions, Economy and Society 8, 79–92. https://doi.org/10.1093/cjres/rsu013

Grable, S.W., 1979. Applying Urban History to City Planning: A Case Study in Atlanta. The Public Historian 1, 45–59. https://doi.org/10.2307/3377281

Gudmundsson, A., Mohajeri, N., 2013. Entropy and order in urban street networks. Scientific Reports 3. https://doi.org/10.1038/srep03324

Haine, A., 2013. Dubai Jumeirah Village Circle: will it be unbroken? The National. https://www.thenational.ae/uae/dubai-jumeirah-village-circle-will-it-be-unbroken-1.469833

Hajrasouliha, A., Yin, L., 2015. The impact of street network connectivity on pedestrian volume. Urban Studies 52, 2483–2497. https://doi.org/10.1177/0042098014544763

Haklay, M., 2010. How Good is Volunteered Geographical Information? A Comparative Study of OpenStreetMap and Ordnance Survey Datasets. Environment and Planning B: Planning and Design 37, 682–703. https://doi.org/10.1068/b35097

Hall, P., 1996. Cities of Tomorrow: An Intellectual History of Urban Planning and Design in the Twentieth Century, 2nd ed. Blackwell Publishers, Malden, MA.

Hanson, J., 1989. Order and Structure in Urban Design. Ekistics 56, 22–42.

Hayden, D., 2004. Building Suburbia: Green Fields and Urban Growth, 1820-2000. Vintage Books, New York, NY.

Holloway, M., 2013. The Measure of Manhattan: The Tumultuous Career and Surprising Legacy of John Randel, Jr., Cartographer, Surveyor, Inventor. W. W. Norton & Company, New York, NY.

Holston, J., 1989. The Modernist City: An Anthropological Critique of Brasília. University of Chicago Press, Chicago, IL.

Hwang, J.-E., Koile, K., 2005. Heuristic Nolli map: a preliminary study in representing the public domain in urban space. Proceedings of the 9th International Conference on Computers in Urban Planning and Urban Management. Presented at CUPUM 2005,

London, England.

Isaac, H., 2014. 10 Bizarre Buildings And Their Fascinating Histories. Gizmodo. https://gizmodo.com/10-bizarre-buildings-and-their-fascinating-histories-1668412190

Ismagilova, E., Hughes, L., Dwivedi, Y.K., Raman, K.R., 2019. Smart cities: Advances in research—An information systems perspective. International Journal of Information Management 47, 88–100. https://doi.org/10.1016/j.ijinfomgt.2019.01.004

Jackson, K.T., 1985. Crabgrass Frontier: The Suburbanization of the United States. Oxford University Press, New York, NY.

Jacobs, A.B., 1984. Looking at Cities. Places 1, 28–37. https://doi.org/10.4159/harvard.9780674863873

Jacobs, A.B., 1995. Great Streets. MIT Press, Cambridge, MA.

Jacobs, J., 1961. The Death and Life of Great American Cities, 1992 ed. Vintage Books, New York, NY.

Jokar Arsanjani, J., Zipf, A., Mooney, P., Helbich, M. (Eds.), 2015. OpenStreetMap in GIScience, Lecture Notes in Geoinformation and Cartography. Springer International, Cham, Switzerland.

Kitchin, R., 2016. The Ethics of Smart Cities and Urban Science. Philosophical Transactions of the Royal Society A: Mathematical, Physical and Engineering Sciences 374, 20160115. https://doi.org/10.1098/rsta.2016.0115

Koeppel, G., 2015. City on a Grid: How New York Became New York. Da Capo Press, Boston, MA.

Kostof, S., 1991. The City Shaped: Urban Patterns and Meanings Through History. Bulfinch Press, New York, NY.

Krivý, M., 2018. Towards a Critique of Cybernetic Urbanism: The Smart City and the Society of Control. Planning Theory 17, 8–30. https://doi.org/10.1177/1473095216645631

Kruse, K.M., 2007. White Flight: Atlanta and the Making of Modern Conservatism. Princeton University Press, Princeton, NJ.

Kubat, A.S., Guney, Y.I., Ozer, O., Topcu, M., Bayraktar, S., 2009. The Effects of the New Development Projects on the Urban Macroform of Dubai: A Syntactic Evaluation, in: Proceedings of the 7th International Space Syntax Symposium. Stockholm, Sweden.

Levinson, D., El-Geneidy, A., 2009. The minimum circuity frontier and the journey to work. Regional Science and Urban Economics 39, 732–738. https://doi.org/10.1016/j.regsciurbeco.2009.07.003

Lin, J., Ban, Y., 2013. Complex Network Topology of Transportation Systems. Transport Reviews 33, 658–685. https://doi.org/10.1080/01441647.2013.848955

Louf, R., Barthelemy, M., 2014. A Typology of Street Patterns. Journal of The Royal Society Interface 11, 1–7. https://doi.org/10.1098/rsif.2014.0924

Lytras, M., Visvizi, A., 2018. Who Uses Smart City Services and What to Make of It: Toward Interdisciplinary Smart Cities Research. Sustainability 10, 1998.

https://doi.org/10.3390/su10061998

Marshall, S., Gil, J., Kropf, K., Tomko, M., Figueiredo, L., 2018. Street Network Studies: from Networks to Models and their Representations. Networks and Spatial Economics. https://doi.org/10.1007/s11067-018-9427-9

Marshall, W., Garrick, N., Marshall, S., 2015. Street Networks, in: International Handbook on Transport and Development. Edward Elgar, Cheltenham, England.

Marshall, W., Piatkowski, D., Garrick, N., 2014. Community design, street networks, and public health. Journal of Transport & Health 1, 326–340. https://doi.org/10.1016/j.jth.2014.06.002

Masucci, A.P., Stanilov, K., Batty, M., 2013. Limited Urban Growth: London's Street Network Dynamics since the 18th Century. PLoS ONE 8, e69469. https://doi.org/10.1371/journal.pone.0069469

Mesh, A., 2014. Feb. 4, 1974: Portland kills the Mount Hood Freeway. Willamette Week. https://www.wweek.com/portland/article-23466-feb-4-1974-portland-kills-the-mount-hood-freeway.html

Micaud, E.C., 1978. Urbanization, urbanism, and the medina of Tunis. International Journal of Middle East Studies 9, 431–447. https://doi.org/10.1017/S0020743800030634

Mohajeri, N., French, J., Gudmundsson, A., 2013a. Entropy Measures of Street-Network Dispersion: Analysis of Coastal Cities in Brazil and Britain. Entropy 15, 3340–3360. https://doi.org/10.3390/e15093340

Mohajeri, N., French, J.R., Batty, M., 2013b. Evolution and entropy in the organization of urban street patterns. Annals of GIS 19, 1–16. https://doi.org/10.1080/19475683.2012.758175

Mohajeri, N., Gudmundsson, A., 2014. The Evolution and Complexity of Urban Street Networks: Urban Street Networks. Geographical Analysis 46, 345–367. https://doi.org/10.1111/gean.12061

Mohajeri, N., Gudmundsson, A., 2012. Entropies and Scaling Exponents of Street and Fracture Networks. Entropy 14, 800–833. https://doi.org/10.3390/e14040800

Moudon, A.V., 1997. Urban morphology as an emerging interdisciplinary field. Urban Morphology 1, 3–10.

Neis, P., Zielstra, D., Zipf, A., 2011. The Street Network Evolution of Crowdsourced Maps: OpenStreetMap in Germany 2007–2011. Future Internet 4, 1–21. https://doi.org/10.3390/fi4010001

Over, M., Schilling, A., Neubauer, S., Zipf, A., 2010. Generating web-based 3D City Models from OpenStreetMap: The current situation in Germany. Computers, Environment and Urban Systems 34, 496–507. https://doi.org/10.1016/j.compenvurbsys.2010.05.001

Porta, S., Latora, V., Wang, F., Rueda, S., Strano, E., Scellato, S., Cardillo, A., Belli, E., Càrdenas, F., Cormenzana, B., Latora, L., 2012. Street Centrality and the Location of Economic Activities in Barcelona. Urban Studies 49, 1471–1488.

    https://doi.org/10.1177/0042098011422570
Robinson, J.B., 2008. Crime and regeneration in urban communities: The case of the big dig in Boston, Massachusetts. Built Environment 34, 46–61. https://doi.org/10.2148/benv.34.1.46
Rose, M., 2001. Atlanta: Then and Now. Thunder Bay Press, San Diego, CA.
Rose-Redwood, R., Bigon, L., 2018. Gridded Worlds: An Urban Anthology. Springer, Cham, Switzerland.
Shanken, A., 2018. The Visual Culture of Planning. Journal of Planning History 17, 300–319. https://doi.org/10.1177/1538513218775122
Shen, Q., Zeng, W., Ye, Y., Arisona, S.M., Schubiger, S., Burkhard, R., Qu, H., 2018. StreetVizor: Visual Exploration of Human-Scale Urban Forms Based on Street Views. IEEE Trans. Visual. Comput. Graphics 24, 1004–1013. https://doi.org/10.1109/TVCG.2017.2744159
Shorto, R., 2004. The Streets Where History Lives. The New York Times, February 9. https://www.nytimes.com/2004/02/09/opinion/the-streets-where-history-lives.html
Smith, M.E., 2007. Form and Meaning in the Earliest Cities: A New Approach to Ancient Urban Planning. Journal of Planning History 6, 3–47. https://doi.org/10.1177/1538513206293713
Söderström, O., 1996. Paper Cities: Visual Thinking in Urban Planning. Ecumene 3, 249–281. https://doi.org/10.1177/147447409600300301
Southworth, M., Ben-Joseph, E., 1997. Streets and the Shaping of Towns and Cities. McGraw-Hill, New York, NY.
Southworth, M., Ben-Joseph, E., 1995. Street Standards and the Shaping of Suburbia. Journal of the American Planning Association 61, 65–81.
Speck, J., 2012. Walkable City: How Downtown Can Save America, One Step at a Time. Farrar, Straus and Giroux, New York, NY.
Stone, M., Knapper, J., Evans, G., Aravopoulou, E., 2018. Information management in the smart city. The Bottom Line 31, 234–249. https://doi.org/10.1108/BL-07-2018-0033
Taylor, R., Rinne, K.W., Kostof, S., 2016. Rome: An Urban History from Antiquity to the Present. Cambridge University Press, Cambridge, England.
Tobler, W.R., 1976. Analytical Cartography. The American Cartographer 3, 21–31. https://doi.org/10.1559/152304076784080230
Townsend, A., 2015. Cities of Data: Examining the New Urban Science. Public Culture 27, 201–212. https://doi.org/10.1215/08992363-2841808
Trudeau, R.J., 1994. Introduction to Graph Theory, 2nd ed. Dover Publications, New York, NY.
Vale, L., 2008. Architecture, Power and National Identity, 2nd ed. Routledge, London.
Vanegas, C.A., Aliaga, D.G., Benes, B., Waddell, P., 2009. Visualization of Simulated Urban Spaces: Inferring Parameterized Generation of Streets, Parcels, and Aerial Imagery. IEEE Trans. Visual. Comput. Graphics 15, 424–435.

https://doi.org/10.1109/TVCG.2008.193

Verstegen, I., Ceen, A., 2013. Giambattista Nolli and Rome: Mapping the City Before and After the Pianta Grande. Studium Urbis, Rome, Italy.

Visvizi, A., Lytras, M.D., Damiani, E., Mathkour, H., 2018. Policy making for smart cities: innovation and social inclusive economic growth for sustainability. Jnl of Science & Tech Policy Mgmt 9, 126–133. https://doi.org/10.1108/JSTPM-07-2018-079

Waddell, P., Garcia-Dorado, I., Maurer, S.M., Boeing, G., Gardner, M., Porter, E., Aliaga, D., 2018. Architecture for Modular Microsimulation of Real Estate Markets and Transportation. Presented at the Applied Urban Modelling Symposium, June 27-29, 2018, Cambridge, England.

Watson, V., 2015. The Allure of "Smart City" Rhetoric: India and Africa. Dialogues in Human Geography 5, 36–39. https://doi.org/10.1177/2043820614565868

Xiao, Y., Webster, C., Orford, S., 2016. Identifying house price effects of changes in urban street configuration: An empirical study in Nanjing, China. Urban Studies 53, 112–131. https://doi.org/10.1177/0042098014560500

Yakunicheva, K., 2014. The Urban Landscape: Perspectives of Structural Development. Presented at the Landscape Transformations Conference, Prague, Czech Republic.

Zhong, C., Arisona, S.M., Huang, X., Batty, M., Schmitt, G., 2014. Detecting the dynamics of urban structure through spatial network analysis. International Journal of Geographical Information Science 28, 2178–2199. https://doi.org/10.1080/13658816.2014.914521

Zielstra, D., Hochmair, H.H., Neis, P., 2013. Assessing the Effect of Data Imports on the Completeness of OpenStreetMap – A United States Case Study. Transactions in GIS 17, 315–334. https://doi.org/10.1111/tgis.12037

Zook, M., Graham, M., Shelton, T., Gorman, S., 2010. Volunteered Geographic Information and Crowdsourcing Disaster Relief: A Case Study of the Haitian Earthquake. World Medical & Health Policy 2, 6–32. https://doi.org/10.2202/1948-4682.1069